\documentclass{aa}
\usepackage{graphics,epsfig,amsmath,amssymb,amstext,txfonts}

\begin{document}

\title{Interactions of UHE cosmic ray nuclei with radiation during acceleration: consequences for the spectrum and composition}

\author{D. Allard \inst{1}   \and  R.J. Protheroe\inst{2}}

\institute{Laboratoire Astroparticule et Cosmologie (APC),
Universit\'e Paris 7/CNRS, 10 rue A. Domon et L. Duquet, 75205
Paris Cedex 13, France.  \and Department of Physics, The
University of Adelaide, SA 5005, Australia. }

\offprints{denis.allard@apc.univ-paris7.fr \\ rprother@physics.adelaide.edu.au}  

\date{Received date; accepted date} 

\abstract{

In this paper, we study the diffusive shock acceleration of
cosmic-ray protons and nuclei, taking all the relevant
interaction processes with photon backgrounds into account. We
investigate how the competition between protons and nuclei is
modified by such acceleration parameters as the acceleration
rate, its rigidity dependence, the photon density, and the
confinement capability of the sources.  We find that protons are
likely to be accelerated to higher energies than nuclei in the
case of interaction-limited acceleration processes, whereas
nuclei are accelerated to higher energies than protons for
confinement-limited acceleration.  Finally, we discuss our
results in the context of possible astrophysical accelerators and
in the light of recent cosmic-ray data.

}

\authorrunning{D. Allard and R. J. Protheroe}

\titlerunning{Interactions of UHE cosmic ray nuclei with
radiation during acceleration}

   \maketitle
%

\section{Introduction}

By plotting magnetic field vs.\ size of various astrophysical
objects, Hillas (\cite{Hillas84}) identified possible sites of
acceleration of UHE CR based on whether or not the putative
source could contain the gyro-radius of the accelerated
particles, and on the likely velocity of scattering centers in
these sites, and narrowed the field of possible sources to radio
galaxy lobes and galaxy clusters.  This issue has been
re-examined recently in the context of diffusive shock
acceleration (DSA) (Protheroe \cite{Protheroe2004}) taking
account of energy losses (synchrotron) and interactions
(Bethe-Heitler and pion photoproduction) which can cut the
spectrum off, and so provide additional constraints.  It was
found that with the maximum possible acceleration rate for a
given magnetic field the spectrum is cut off by pion
photoproduction if the magnetic field is below $\sim$3~$\mu$G,
and for higher fields the spectrum is cut off by synchrotron
losses.

However, for an acceleration rate $10^{-4}$ times the
maximum possible, pion photoproduction cuts off the spectrum for
fields below $\sim$100~$\mu$G and synchrotron losses cut off the
spectrum for higher fields.  In order to accelerate protons above
$10^{20}$eV the ideal sources were found to have acceleration
region sizes (magnetic field) between 1~kpc ($\sim$1~mG) and
1~Mpc ($\sim$1~$\mu$G) such that the gyro-radius can be contained
within the accelerator.  These parameters allow the spectrum to
extent beyond $10^{20}$eV where it will be cut off either by
synchrotron losses (for the higher magnetic fields in this range)
or by pion photoproduction (for the lower fields in this range).
This makes radio galaxies and clusters of galaxies the probable
sites of UHE CR acceleration (see Fig.~6 of Protheroe
\cite{Protheroe2004}) in agreement with the conclusion of Hillas.
However, if the whole acceleration region is moving
relativistically, Doppler boosting in energy of neutrons
(produced by pion photoproduction) escaping from such sources
allows blazar jets (Protheroe et
al.~\cite{ProtheroeDoneaReimer03}) and GRB (Waxman
\cite{Waxman2006} and references therein) to remain possible
sites of UHE CR acceleration, and in these cases proton
synchrotron losses are likely to cut off the spectrum.
Large-scale cosmic shocks in the Universe may also be sites of
UHE CR acceleration (Kang \& Jones \cite{KangJones02}, Inoue et
al.\ \cite{Inoue_etal2007}) as well as colossal
magneto-hydrodynamic fossil AGN structures (Benford \& Protheroe
\cite{BenfordProtheroe08}), AGN jets and cocoons surrounding AGN
jets (Schopper et al.\ \cite{Schopperetal02}, Norman et al.\
\cite{Normanetal95}) and hot spots in radio lobes of powerful
radio galaxies (Rachen \& Biermann \cite{RachenBiermann93}).  In
these cases pion photoproduction, the size of the accelerator or
the time available for acceleration may cause the cut-off in the
spectrum of protons at acceleration.  In a recent review,
Biermann et al.\ (\cite{Biermann_etal2008}) concluded that the
most promising contenders are radio galaxies and GRB.

Apart from the nature of the cosmic accelerators, the composition
of the accelerated particles at the highest energies is a key
parameter in understanding the origin of the extragalactic
cosmic-rays and the physical conditions at play at the sources.
To determine the spectrum and composition of accelerated
particles at an astrophysical source the injected composition at
the shock and the interactions within the acceleration region
should be considered.  Satellite measurements of low energy
galactic cosmic-rays show that the cosmic-ray composition is
deficient in H with respect to interstellar matter, and the
relative abundances of the various species are believed to be
related to their volatility (Meyer et al., \cite{Meyer}). It is
of course possible that such enrichment takes place also in
extragalactic sources.

In the present paper, we take into account all the relevant
interactions of protons and nuclei with ambient photon
backgrounds and calculate the spectrum and composition resulting
from diffusive shock acceleration for various compositions
injected at the shock.  Using a box model of DSA (Protheroe
\cite{Protheroe2004}), the SOPHIA event generator for pion
photoproduction (M\"ucke et al., \cite{Muecke00}) and the
photon-nucleus interaction event generator developed in Allard et
al. (\cite{Denis2005a}-\cite{Denis2008}), we study the dependence
of the composition and spectrum of accelerated cosmic rays on the
acceleration rate, the rigidity dependence of the diffusion
coefficient, the confinement within the acceleration site, and
the density of the photon backgrounds.  The paper is organized as
follows, in the next section we introduce the relevant
interaction processes of protons and nuclei with photon
backgrounds. In Sect.~3 we describe the box model for DSA we use
in our calculations. In Sect.~4 we present the spectra and
composition resulting from the acceleration processes under
different hypotheses about the injected composition and physical
parameters for the accelerators. Finally, in Sect.~5 we discuss
our results.


\section{Interactions of nucleons and nuclei with radiation and their modeling}
\label{Interactions}

Protons and nuclei propagating through the intergalactic medium
interact mainly with cosmic microwave background (CMB),
infrared, optical and ultraviolet (IR/Opt/UV) photon
backgrounds. These interactions produce features in the
propagated UHE CR spectrum such as the ``GZK cutoff'' (Greisen,
\cite{GZK66}; Zatsepin \& Kuzmin, \cite{GZK66b}).
Pions produced in the same interactions generate the cosmogenic neutrino flux (Berezinsky \&
Zatsepin, \cite{BereOriginal}). During the acceleration process,
these interactions are also expected to take place at a rate
depending on the ambient photon density and can also produce
spectral and composition features in the cosmic-rays as well as
secondary particles (neutrinos and photons), as discussed by
Protheroe (\cite{Protheroe2004}) in detail for the case of proton
primaries. 

\begin{figure}
\centering
   \includegraphics[height=8cm,width=8cm]{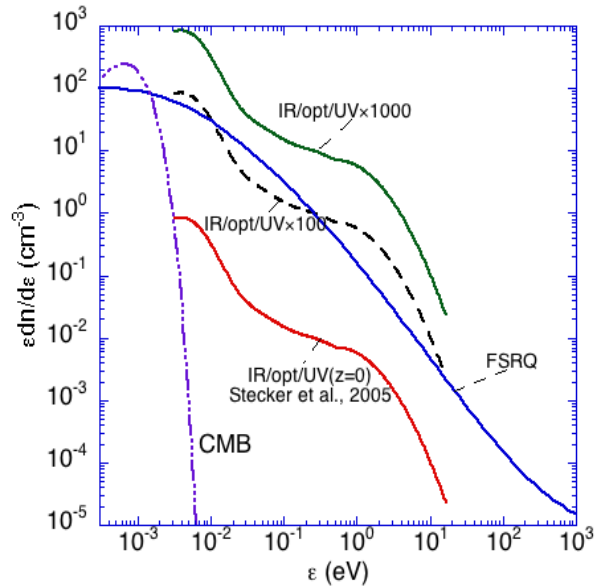}
     \caption{Radiation field of the CMB compared with the
IR/Opt/UV from Stecker et al.\ (\cite{MS05}) with various scaling
factors.  Also shown, as will be discussed in Sect.~4, is the
radiation field expected at 200~kpc along the jet from the
flat-spectrum radio quasar (FSRQ) 0208-512 based on its spectral
energy distribution given in fig.~5 of Tavecchio et al.\
(\cite{Tavecchio2002}).}
     \label{FSRQ}
\end{figure}

In the following, we will 
use the recent estimate of Stecker et al. (\cite{MS05}) for the
intergalactic background to model the photon backgrounds in
the infra-red, optical and ultra-violet window.  We re-scale its density with different
scaling factors to account for the higher density IR/opt/UV
backgrounds that could be expected in astrophysical sources (see
Fig.~\ref{FSRQ}).

\begin{figure*}[t]
\centering
\hfill\includegraphics[height=8cm]{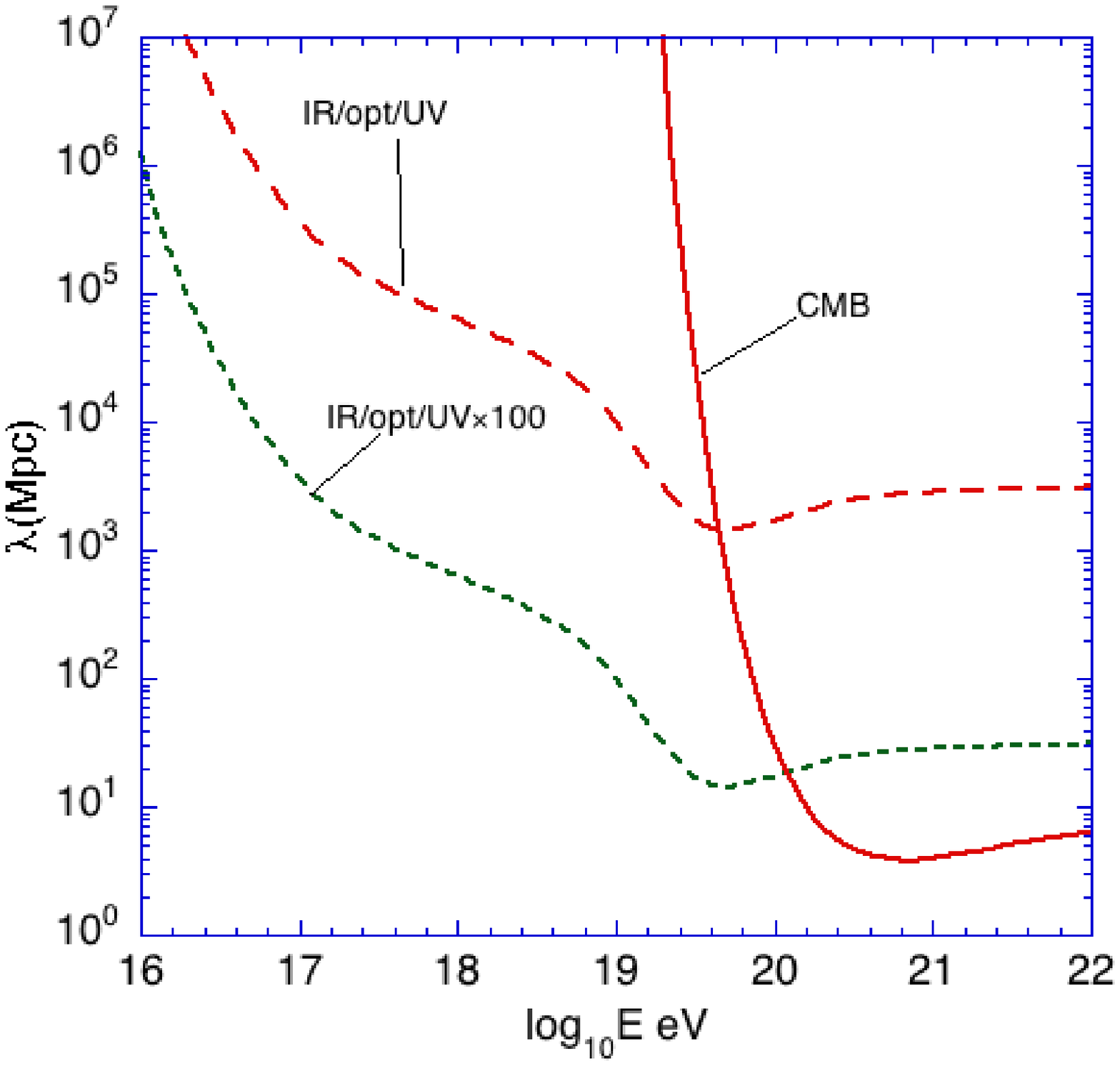}\hfill
\includegraphics[height=8cm]{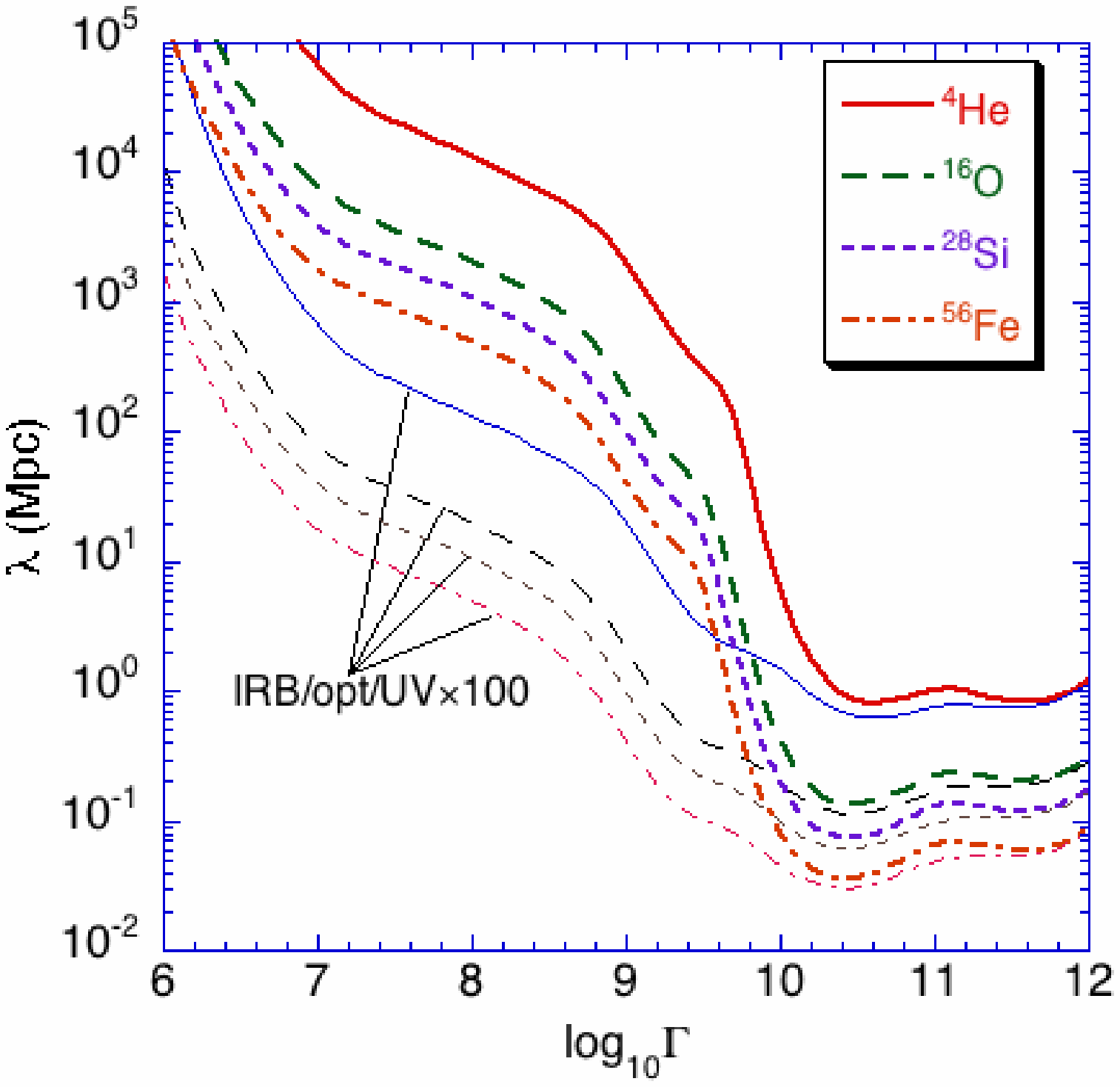}\hfill~
\caption{(a) Left: Energy dependence of the mean free path for
pion photoproduction of cosmic-ray protons in the intergalactic
background radiation fields, and for the case where the IR/Opt/UV
background is 100 times higher. (b) Right: Lorentz-factor
dependence of the mean free path for photo-disintegration for four
different nuclei (He, O, Si, Fe) in the same photon backgrounds
as in the left panel.}
\label{MFP}
\end{figure*}

Bethe-Heitler pair production interactions with the various
photon backgrounds are treated as a continuous energy loss
process. We use the cross sections, inelasticities and the mass
and charge scaling of the attenuation length for nuclei provided
by Rachen (\cite{Rachen}).  If the energy of the background
photons exceeds $\sim$145 MeV in the nucleon rest frame, protons
and neutrons can interact through the pion photoproduction
process.  This high inelasticity process is better treated by
Monte Carlo calculation, and our Monte Carlo code uses the SOPHIA
event generator (M\"{u}cke et al., 2000) allowing us to treat
accurately the various interaction channels (direct pion
production, resonances, multi-pion production) and their
branching ratios (see Rachen \cite{Rachen}).  The energy
dependence of the mean free path of protons for photopion
production with CMB and IR/opt/UV photons is shown in
Fig.~\ref{MFP}a.  Also shown is the mean free path for a
IR/opt/UV background 100 times denser.

The interactions experienced by nuclei with photon backgrounds
differ from those of protons.  In addition to the pair production
losses that result in a decrease of the Lorentz factor and the
rigidity of the nucleus, one must consider the
photo-disintegration (also called photo-erosion) processes that
leads to the ejection of one or several nucleons from the
nucleus. Different photo-erosion processes dominate the total
interaction cross section at different energies (Puget et al.,
\cite{PSB}).  The lowest energy disintegration process is the
Giant Dipole Resonance (GDR) which results in the emission of one
or two nucleons and $\alpha$ particles.  The GDR process is the
most relevant as it has the highest cross section and the lowest
thresholds, between 10 and 20 MeV, for all nuclei.  For nuclei
with mass $A \geq 9$, we use the theoretical GDR cross sections
presented by Khan et al. (\cite{Khan04}), which take into account
all the individual reaction channels ($n$, $p$, 2$n$, 2$p$, $np$,
$\alpha$, \dots) and are in better agreement with data than
previous treatments.  For nuclei with $A < 9$, we use the
phenomenological fits to the data provided by Rachen
(\cite{Rachen}).  Between $\sim$30 MeV in the nucleus rest frame
and the photopion production threshold the quasi-deuteron (QD)
process becomes comparable to the GDR, and dominates the total
cross section at higher energies.

Pion photoproduction (baryonic resonance (BR) process) by nuclei
becomes relevant above 150 MeV in the nucleus rest frame, i.e.,
$\sim 5\times 10^{21}$ eV in the lab frame for iron nuclei
interacting with the CMB.  We use the parameterization given by
Rachen (\cite{Rachen}) in which the cross section in this energy
range is proportional to the mass of the nucleus.  The basis for
this scaling that of the deuteron photoabsorption cross section
which is known in great detail, although nuclear shadowing
effects are expected to break this scaling above 1 GeV.  It is
important to note that pion photoproduction cross sections for
nuclei are different from the free nucleon case. In particular,
in nuclei the baryonic resonances heavier than the first $\Delta$
resonance are far less pronounced than for nucleons, and the
cross sections can not be simply derived from the free nucleon
case. We also follow Rachen (\cite{Rachen}) for the treatment of
secondary particle production, nucleon multiplicities, energy
losses, and the probability of absorption of the produced pion by
the parent nucleus.

The Lorentz-factor dependence of the total mean free path (all
photo-disintegration processes and all channels) for
photo-disintegration for four different abundant species is shown
in Fig.~\ref{MFP}b for the same photon backgrounds as
Fig.~\ref{MFP}a.  One can see that the interaction thresholds
occur at similar Lorentz factors for all the species, and that
the mean free paths scale approximately linearly with the mass as
a consequence of the mass scaling of the cross sections.
Photo-disintegration is dominated by the GDR process over the
whole energy range, except for the second minimum at the highest
energies which is due to the BR process (see Fig.~3b of Allard et
al. (\cite{Allard2006}) for the contributions of the various
processes to the mean free path).

\section{Diffusive shock acceleration}

Diffusion of energetic particles in magnetic fields depends on
their magnetic rigidity $\rho\equiv pc/Ze$, where $p$ is the the
particle's momentum and $Ze$ its charge.  In DSA the rigidity
gain and escape rates depend on the diffusion coefficient which
is usually assumed to have a power-law dependence on rigidity,
$\kappa(\rho) \propto \rho^\delta$, with the exponent depending
on the nature of the turbulence present in the magnetic field:
$\delta=1/3$ (Kolmogorov spectrum), 1/2 (Kraichnan spectrum) or 1
(completely disordered field).  The diffusion coefficient is
lowest, and the acceleration fastest, for the case where the
magnetic field $B$ is highly disordered, and it can approach the
Bohm diffusion coefficient, $\kappa_B={1 \over 3} R_g c$ where
$R_g=\rho/cB$ is the gyro-radius.  In this case Jokipii
(\cite{Jokipii87}) showed that if the diffusion coefficient
parallel to the magnetic field direction is a factor $\eta > 1$
times the Bohm limit, i.e.\
\begin{equation}
\kappa_\parallel = \eta {1 \over 3} r_g c
\end{equation}
then the diffusion coefficient perpendicular to the
magnetic field  is expected to be approximately
\begin{equation}
\kappa_\perp \approx {k_\parallel \over 1 + \eta^2}
\end{equation}
provided that $\eta$ is not too large (values in the range up to
10 appear appropriate).

Under this assumption, and for the case of DSA at a strong plane
parallel shock (i.e.\ with magnetic field $B$ parallel to the
shock normal) with shock speed $u_1$, this leads to a rigidity
gain rate
\begin{equation}
r_{\rm gain}(\rho) \equiv {1 \over \rho} \left. {d\rho \over dt} \right|_{\rm gain} \approx {3 u_1^2 B \over 20 \eta}\rho^{-1},
\end{equation}
and for a perpendicular shock (i.e. with magnetic field
perpendicular to the shock normal) the rigidity gain rate would
be a factor $\sim 2.5\eta(1+\eta^2)$ higher (Protheroe
\cite{Protheroe_BookChapter1998}).

In general, for particle acceleration by electric fields induced by the
motion of magnetic fields $B$ (including those at astrophysical
shocks), the maximum  rigidity gain rate of relativistic
particles can be written
\begin{equation}
r_{\rm gain}  = \xi(\rho)  c^2 B \rho^{-1}.
\label{eq:max_acc}
\end{equation}
where $\xi(\rho)$$<$1 is the acceleration rate parameter which
depends on the details of the acceleration mechanism, with
$\xi$=1 corresponding to the highest posible gain rate for any
mechanism.  For DSA $\xi(\rho)$ will depend on the shock
velocity, magnetic field alignment and diffusion coefficients.
For example, for the parallel shock case with $u_1=0.1c$ and
$\eta=10$ one finds $\xi(\rho) \leq 1.5\times 10^{-4}$, and
for the equivalent perpendicular shock case $\xi(\rho) \leq
0.04$ (see Protheroe \cite{Protheroe_BookChapter1998}).  Detailed
and rigorous treatments of DSA are given in several review
articles (Drury \cite{Drury83a}, Blandford \& Eichler
\cite{BlandfordEichler87}, Berezhko \& Krymsky
\cite{BerezhkoKrymsky88}); see particularly the review by Jones
\& Ellison (\cite{JonesEllison91}), on the plasma physics of
shock acceleration, which also includes a brief historical review
and refers to early work.

Protheroe and Stanev (\cite{ProtheroeStanev99}) proposed a box
model of DSA (based on that developed by Szabo \& Protheroe
\cite{Szabo94}) which was able to reproduce the essential
features of DSA in a simple leaky-box scenario.  Particles were
injected into the ``box'', and while inside the box their energy
increased at the {\em average} rate that would apply given the
physical parameters of the shock acceleration being modelled,
i.e.\ the upstream and downstream diffusion coefficients, the
shock velocity and compression ratio.  Particles being
accelerated were made to leak out of the box, again at the {\em
average} rate that would apply given the physical parameters, and
these particles represented the accelerated particles escaping
downstream in the standard shock acceleration picture.  Protheroe
(\cite{Protheroe2004}) incorporated improvements suggested by
Drury et al.\ (\cite{Druryetal99}) into the model, and used it to
investigate cut-offs and pile-ups in spectra of accelerated
particles in the presence of energy losses or interactions,
including the case of acceleration and propagation of UHE cosmic
ray protons in the presence of the CMB.  It was found that
although almost all information about the acceleration conditions
was lost after propagation, as a result of the GZK cut off,
significant information remained imprinted in the spectrum of
cosmogenic neutrinos produced in the ``GZK interactions''.  In
the present paper we use the calculation scheme of Protheroe
(\cite{Protheroe2004}) as it is very convenient to implement in a
numerical or Monte Carlo program to investigate the spectrum and
composition of UHE cosmic rays accelerated in the presence of the
CMB and IR/Opt/UV radiation fields.
\begin{figure}[b]
\centering{\includegraphics[height=7cm]{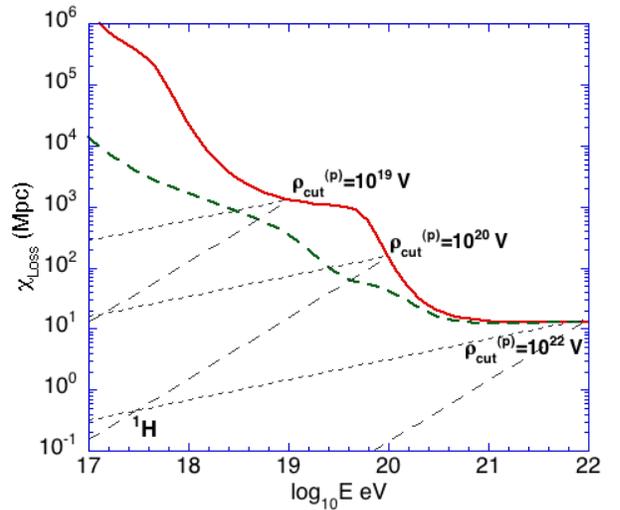}}
\caption{Energy dependence of the loss length for cosmic-ray
protons in the CMB plus intergalactic IR/Opt/UV background (continuous
line) and assuming a IR/Opt/UV background 100 times higher
(dashed line). The loss length is compared to the gain length
assuming three different values for $\rho_{cut}^{(p)}$
($10^{19}$, $10^{20}$ and $10^{22}$ V) and $\delta=1/3$ (thin
short dashed lines) and $\delta=1$ (thin long dashed lines). }
\label{Loss1}
\end{figure}

As we have seen, the normalization of the gain rate depends in a
non-trivial way on the magnetic field, turbulence spectrum, shock
configuration and shock velocity.  Different classes of source,
and even different sources of the same class, will accelerate
particles up to different maximum energies.  We have therefore
arbitrarily chosen to normalize the gain rate at a given cut-off
rigidity to that necessary for protons to reach that given
cut-off rigidity in a steady state situation in the presence of
the CMB and IR/Opt/UV radiation.  We have done this irrespective
of the rigidity dependence of $r_{\rm gain}(\rho)$ for two
reasons: firstly because we are interested in the {\em maximum}
energies achieveable and in interactions occurring at or near
cut-off, and secondly because it is only the gain rate {\em close
to the cut-off rigidity}, not the rigidity-dependence of the gain
rate, that determines the value of the cut-off rigidity, at least
for a steady-state situation.

The nominal cut-off rigidity for protons $\rho^{(p)}_{\rm cut}$
occurs where the gain rate equals the loss rate in the
intergalactic IR/Opt/UV background,
\begin{eqnarray}
r_{\rm gain}(\rho^{(p)}_{\rm cut})=r_{\rm loss}(\rho^{(p)}_{\rm cut},Z=1).
\end{eqnarray}
In Fig.~\ref{Loss1} we show the energy dependence of the loss (or
attenuation) length for cosmic-ray protons, $x_{\rm
loss}(\rho,Z=1)$, in the CMB plus intergalactic IR/Opt/UV
background (continuous line) and in the CMB plus an intergalactic
IR/Opt/UV background 100 times higher (dashed line) (assuming the
same photon backgrounds as in Fig.~\ref{MFP}).  We have added to
Fig.~\ref{Loss1} curves representing $x_{\rm gain}(\rho)\equiv
c/r_{\rm gain}(\rho)$ for different values of $\rho_{\rm
cut}^{(p)}$ (assuming $\delta$=1/3 and $\delta$=1).  Using the
value of $\rho_{\rm cut}^{(p)}$ in the CMB plus intergalactic
IR/Opt/UV background to define the normalization of the rigidity
gain rate we have
\begin{eqnarray}
r_{\rm gain}(\rho) = r_{\rm gain}(\rho_{\rm cut}^{(p)})\left({\rho \over \rho_{\rm cut}^{(p)}}\right)^{-\delta} 
= r_{\rm loss}(\rho_{\rm cut}^{(p)},Z=1)\left({\rho \over \rho_{\rm cut}^{(p)}}\right)^{-\delta} .
\label{eq:r_gainEQr_cut}
\end{eqnarray}

\begin{figure}[t]
\centering{\includegraphics[height=7cm]{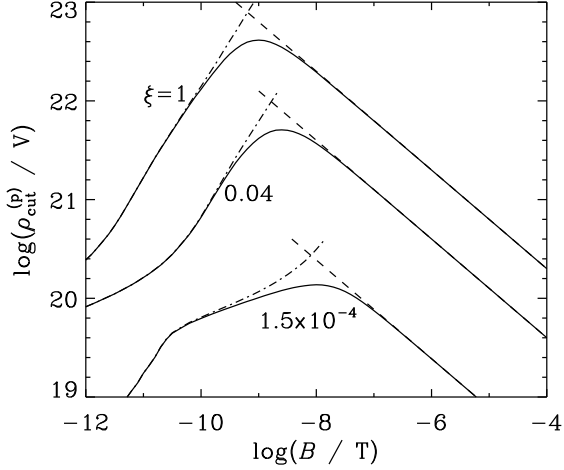}}
\caption{Cut-off rigidity of protons accelerated in the presence of CMB radiation vs.\ magnetic field for $\xi$=1, 0.04 and $1.5\times 10^{-4}$: chain curve - cutoff due to Bethe-Heitler pair production and pion photoproduction; dashed curve - cutoff due to synchrotron; solid curve - total. }
\label{fig_rho_cut_vs_b}
\end{figure}

We have chosen for $\rho_{cut}^{(p)}$ values in the range
$10^{19}$~V to $10^{22}$~V, and we now discuss for what magnetic
fields and acceleration rate parameters it is feasible for
protons to achieve these cut-off rigidities in the CMB plus
IR/Opt/UV radiation.  For $\rho_{cut}^{(p)}=10^{19}$, $10^{20}$,
$10^{22}$ and $10^{22}$~V, the corresponding values of
$\xi(\rho_{cut}^{(p)})B$ are $1.5\times 10^{-15}$, $6.9\times
10^{-14}$ , $6.9\times 10^{-12}$ and $1.6\times 10^{-10}$~T.
However, for a given value of the product $
\xi(\rho_{cut}^{(p)})B$, not all vales of $B$ are possible as
protons will lose energy also by synchrotron radiation, and this
actually becomes more important than pion photoproduction losses
at high rigidity.  If the proton spectrum is cut-off by
synchrotron loss, then the cut-off rigidity is given by
\begin{eqnarray}
\rho^{(p)}_{\rm cut}=2\times 10^{18}\xi^{1/2}B^{-1/2} \mbox{~~V}
\end{eqnarray}
where $B$ is in tesla.  To illustrate this, we plot in Fig.~\ref{fig_rho_cut_vs_b} the cut-off rigidity vs.\ magnetic field taking into account of synchrotron, Bethe-Heitler pair production and pion photoproduction losses for three values of acceleration rate parameter.
For example, for the parallel shock case with $u_1=0.1c$ and $\eta=10$ for which $\xi(\rho) \approx 1.5\times 10^{-4}$ it is possible for protons  to reach rigidities above $\rho =10^{19}$~V for magnetic fields in the range $5\times 10^{-12}$ to $5\times 10^{-6}$~T, and to reach rigidities above $\rho =10^{20}$~V fields in the range $10^{-9}$ to $5\times 10^{-8}$~T are required with  the maximum rigidity possible for that case being $1.5\times 10^{20}$~V achievable for $10^{-8}$~T.
For the equivalent perpendicular shock case for which $\xi(\rho) \approx 0.04$, $\rho=10^{21}$~V can be reached for magnetic fields in the range $10^{-10}$ to $2\times 10^{-7}$~T, with the maximum rigidity possible being $5\times 10^{21}$~V achievable for $10^{-9}$~T.  Higher rigidities require relativistic shocks or different mechanisms, but  proton rigidities above $4\times 10^{22}$~V are are not possible without Doppler boosting of neutrons produced in pion photoproduction within a relativistically  moving source such as an AGN jet or GRB.

\begin{figure}[t]
\centering{\includegraphics[height=7cm]{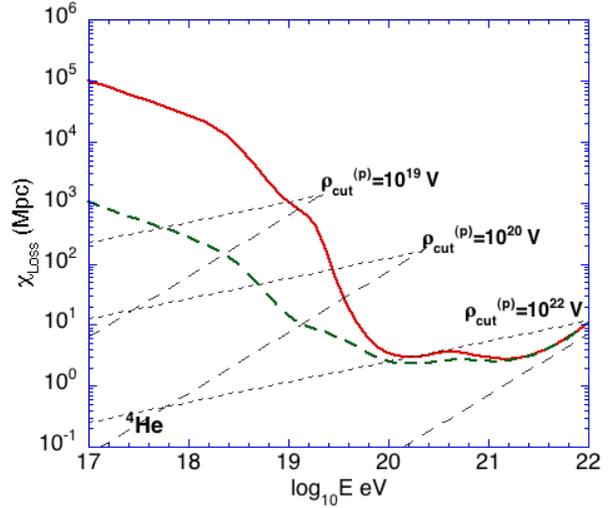}}
\centering{\includegraphics[height=7cm]{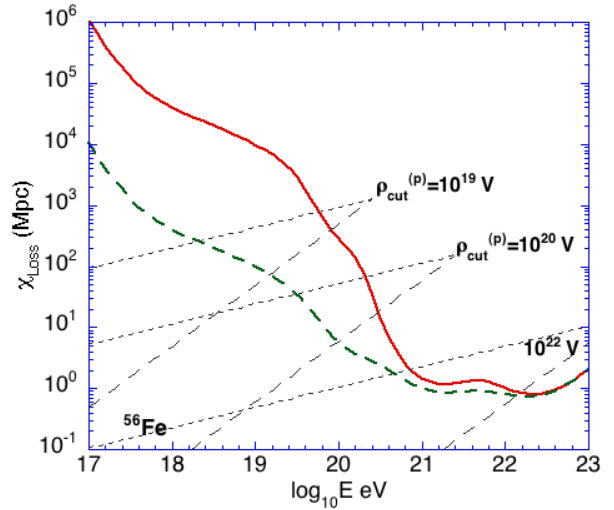}}
\caption{As Fig.~\ref{Loss1} but for (a) He (top), and (b) Fe nuclei (bottom).}
\label{Loss2}
\end{figure}

In Fig.~\ref{Loss2} the energy dependence of the loss length for
He (Fig.~\ref{Loss2}a) and Fe (Fig.~\ref{Loss2}b) is plotted
(assuming the same photon backgrounds as in Fig.~\ref{MFP}) and
is compared to the energy dependence of gain lengths for the same
{\em rigidity dependences} as the curves in Fig.~\ref{Loss1}. As
the abscissa refers to the energy, and the acceleration depends
on rigidity, the curves for the gain length for a given
$\rho_{\rm cut}^{(p)}$ are shifted in energy proportionally to
the charge (i.e. by a factor of 2 for He and a factor of 26 for
Fe).  Although the loss length is not strictly relevant for nuclei
because nuclei do not remain on the same energy loss curve when
losing energy, unlike protons, these figures show that the output
of the acceleration process will crucially depend on the
normalization of the background radiation and the gain rate and
its rigidity dependence, and that the ``competition'' between
protons and nuclei will, at least for some combinations of
parameters, be non-trivial.

\subsection{Box model approximation}

In the box model of shock acceleration, particles of magnetic
rigidity $\rho_0$ are injected into the acceleration zone, or
``box'', and while inside the box are accelerated at a rate
$r_{\rm tot}(\rho)$ and escape from the box at a rate $r_{\rm
esc}(\rho)$.  These two rates uniquely determine the spectrum of
accelerated particles, i.e.\ those escaping from the box.  

The
average rate of change of rigidity of a nucleus of rigidity
$\rho$ and atomic number $Z$ during shock acceleration in the
presence of energy losing processes is
\begin{eqnarray}
\left.{d\rho \over dt}\right|_{\rm total, Z} & \equiv &  \rho r_{\rm tot}(\rho, Z) =
\rho\left[r_{\rm gain}(\rho)- r_{\rm loss}(\rho, Z) \right]
\end{eqnarray} 
where $r_{\rm loss}(\rho, Z)$ includes both losses due to
interactions such as Bethe-Heitler pair production that can be
approximated as continuous energy losses, and pion
photoproduction in which a significant fraction of a proton's
energy is lost per interaction.  Hence, using
Eq.~\ref{eq:r_gainEQr_cut} we have
\begin{eqnarray}
r_{\rm tot}(\rho,Z) = r_{\rm loss}(\rho_{\rm cut}^{(p)},Z\!=\!1)\left({\rho \over \rho_{\rm cut}^{(p)}}\right)^{-\delta} - r_{\rm loss}(\rho,Z).
\label{eq_r_acc}
\end{eqnarray}

If there are no energy losses and no additional escape processes
a power-law spectrum results, and the integral spectral index ,
($\Gamma$-1), equals the ratio of the escape rate to the
gain rate.  In this case, the escape rate representing
escape downstream is $(\Gamma-1)r_{\rm gain}(\rho)$,
and for injection of $N_0$ particles of rigidity $\rho_0$ the expected
differential rigidity spectrum of accelerated particles is
\begin{eqnarray}
\Phi_0(\rho) = N_0 (\Gamma-1)\rho_0^{\Gamma-1}\rho^{-\Gamma}.
\label{Phi0}
\end{eqnarray}
 
The acceleration zone extends distances
$L_1(\rho)=\kappa_1(\rho)/u_1$ and $L_2(\rho)=\kappa_2(\rho)/u_2$
upstream and downstream from the shock, respectively where $u_1$
and $u_2$ are the upstream and downstream flow velocities in the
shock frame.  Adding an extra escape term $r_{\rm esc}^{\rm
max}(\rho,\rho_{\rm max})$ responsible for the ``maximum
rigidity'' for example arising from the finite size of the
accelerator, and an escape term $r_{\rm esc}^{\rm fallout}(\rho)$
resulting from energy losses causing particles to ``fall out of
the box'' gives the total escape rate
\begin{eqnarray}
r_{\rm esc}(\rho) &=&  (\Gamma-1)r_{\rm loss}(\rho_{\rm cut}^{(p)},Z\!=\!1)\left({\rho \over \rho_{\rm cut}^{(p)}}\right)^{-\delta} + \nonumber \\ && \;\;\;\; + \; r_{\rm esc}^{\rm max}(\rho,\rho_{\rm max}) + r_{\rm esc}^{\rm fallout}(\rho,Z). 
\end{eqnarray}
For the case of diffusion (with diffusion coefficient $\kappa
\propto \rho^\delta$) perpendicular to the shock normal to the edge of
the acceleration region being responsible for the maximum rigidity,
\begin{eqnarray}
r_{\rm esc}^{\rm max}(\rho,\rho_{\rm max}) =  (\Gamma-1)r_{\rm loss}(\rho_{\rm cut}^{(p)},Z\!=\!1)\left({\rho_{\rm max}^2 \over \rho \, \rho_{\rm cut}^{(p)}}\right)^{-\delta}.
\label{eq:sideways}
\end{eqnarray}
For the escape rate resulting from energy losses
\begin{eqnarray}
r_{\rm esc}^{\rm fallout}(\rho,Z)= {1 \over L(\rho)} {d L_2 \over d\rho}\left.{d \rho \over dt}\right|_{{\rm loss},Z}  = \delta \,\ell_2\, r_{\rm loss}(\rho,Z),
\end{eqnarray}
where $L=(L_1+L_2)$, and
 we adopt $\ell_2 \approx 0.5$ (Protheroe \cite{Protheroe2004}).

Interactions of relativistic particles with matter or radiation
generally result in the interacting particle losing energy, and
this energy lost is either given to the struck particle or photon
(elastic collision) or used in production of secondary particles
(inelastic collision).  If the mean interaction length (mean free
path) is $\lambda_{\rm int}(\rho)$, for ultra-relativistic
particles the rate of interaction is $r_{\rm
int}(\rho)=c/\lambda_{\rm int}(\rho)$.  If the mean inelasticity,
i.e.\ average fraction of energy lost per interaction, is
$\bar{\alpha}(\rho)$ then the effective loss rate for these
non-continuous losses is $r_{\rm loss}^{\rm
(int)}(\rho)=\bar{\alpha}(\rho)r_{\rm int}(\rho)$.  Of course, if
$\bar{\alpha}(\rho) \ll 1$ then one can approximate the
interactions as continuous energy losses.

In the Monte Carlo simulations Bethe-Heitler pair production was
treated as a continuous loss processes, but hadronic collisions
of protons and neutrons as well as  photo-disintegration of nuclei
were treated by the Monte Carlo method.  Neutrons produced, e.g.\
in $p\gamma \to n\pi^+$ or during photo-disintegration of nuclei 
($N\gamma \to N^\prime + in + jp$, where $N$ and $N^\prime$ are the
parent and daughter nuclei, and $i$ and $j$ are integers), may
decay inside the box into protons which continue to be
accelerated, or they may escape from the box to decay outside
into cosmic rays depending on the neutron's Lorentz factor and
the physical dimensions of the acceleration region.  Since the
simulations are for a given $\rho_{\rm cut}^{(p)}$, $\rho_{\rm max}$ and
$\delta$, and the size of the box depends on the (unknown)
normalization of the diffusion coefficients, we have performed
simulations separately for the two extreme cases.  

In the simulation with $N_0$ particles injected, a particle is
injected at time $t$=0 with rigidity $\rho(0)=\rho_0$ and
statistical weight $w(0)=w_0=1/N_0$.  Its subsequent rigidity,
$\rho(t)$, and weight, $w(t)$, are determined after successive
time steps $\Delta t$ chosen to be much smaller than the smallest
time-scale in the problem.  In each time step, first the rigidity
is changed, $\rho(t+\Delta t)=\rho(t)[1+\Delta t \, r_{\rm
acc}(\rho)]$, and then the probability of escaping in time
$\Delta t$ is estimated as $P_{\rm esc}=\{1-\exp[-\Delta t \,
r_{\rm esc}(\rho)]\}$. Then $w(t)P_{\rm
esc}$ particles with energy $E$ are binned in the histogram of
accelerated particles for the particle's species, and the particle's weight is changed to
reflect the fraction not escaping, $w(t+\Delta t)=w(t)(1-P_{\rm
esc})$.

Next, the probability of interacting in time $\Delta t$ is
estimated, $P_{{\rm int},A,Z}=\{1-\exp[-\Delta t \, r_{\rm
int}(\rho,A,Z)]\}$, and a random number is generated to determine
whether or not an interaction takes place.  If an interaction does take place, the
energy of the target photon in the proton or nucleus rest frame (NRF),
$\epsilon'$, is sampled from
\[
p(\epsilon')\propto n(\epsilon')\sigma(\epsilon') \mbox{~~~~ for
~~} \epsilon_{\rm thr}' < \epsilon' < 2 \gamma \epsilon_{\rm max}
\]
where $n(\epsilon)d\epsilon$ is the LAB-frame number density of
target photons with energy in the range $\epsilon$ to
$(\epsilon+d\epsilon)$, $\sigma(\epsilon')$ is the relevant cross
section in the NRF, $\epsilon_{\rm thr}'$ is the NRF threshold
energy for the process considered, $\gamma$ is the Lorentz factor
of the proton or nucleus, and $\epsilon_{\rm max}$ is maximum
target photon energy in the LAB frame.  For ultra-relativistic
projectiles, the photon's direction in the NRF will be
anti-parallel to the LAB-frame direction of the projectile.  In
the nucleon case (either a primary proton or a nucleon resulting
from photo-disintegration), pion photoproduction interactions are
simulated in the the NRF using the SOPHIA event generator, and
the 4-momenta of all the produced particles are Lorentz
transformed to the LAB frame.  If the leading nucleon remains a
proton, its rigidity ($\rho_i$ before the interaction) is changed
to $\rho = \rho_i(1-\alpha)$, where $\alpha$ is the inelasticity
for that collision.  However, if it is a neutron we consider two
extreme possibilities as discussed earlier: (a) it decays to a
proton inside the acceleration zone and will continue to be
accelerated, or (b) it escapes from the acceleration zone,
decaying outside and adding to the pool of cosmic rays which have
escaped from the accelerator.  Escaping protons, neutrons, nuclei
and neutrinos are binned in energy.

In the case of interactions of nuclei, the number of emitted
nucleons and $\alpha$ particles is determined according to
branching ratios of the various interaction channels.  Whenever
the BR process is chosen, we check whether or not neutrinos are
produced and calculate their energy according to the $\Delta$
resonance kinematics (see Allard et al. \cite{Allard2006} for
more details)

Next we take account of the additional escape resulting from the
rigidity decreasing during the interaction.  This is more
important for pion photoproduction losses than for
photo-disintegration in which the rigidity does not necessarily
decrease. The probability
of the particle immediately escaping downstream due to falling
out of the box is estimated,
\begin{eqnarray}
{\rm Prob.(escape},\rho_i \to \rho) = {L_2(\rho_i)-L_2(\rho) \over
L(\rho_i)} = \left[1-\left({\rho \over
\rho_i}\right)^\delta\right]\ell_2,
\end{eqnarray}
and a random number is generated to determine if this happens.
If the particle does escape, then $w(t+\Delta t)$ particles with
momenta $\rho$ are binned in a histogram of accelerated
particles, and a new particle is injected with rigidity $\rho_0$
and weight $w_0$.  If the particle does not escape, then the
particle's rigidity and weight are evolved through a new time
step as described above.  Finally, if the time since injection
exceeds a maximum time which may represent the time during which
the accelerator is active, $t_{\rm active}$, the acceleration
ceases and a new particle is injected.  In most cases we set to
$t_{\rm active}$=10~Gyr, but we shall explore shorter activity
time scales.

\begin{figure*}
\centering
   \includegraphics[angle=90,height=14cm,width=19cm]{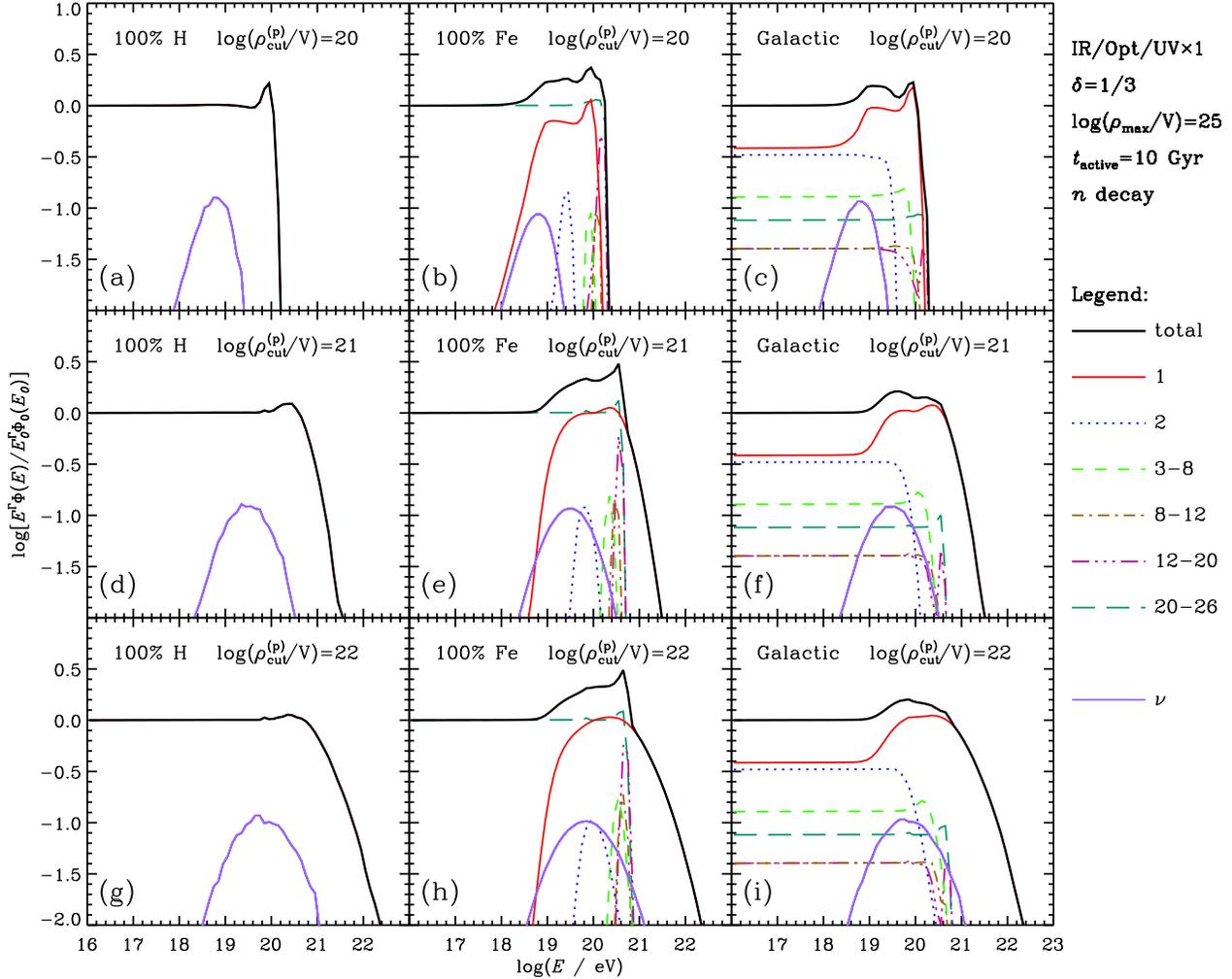}
     \caption{Spectra of the various nuclei emerging after
     acceleration together with the spectra of neutrons and
     neutrinos (sum of all flavors) escaping from the
     acceleration region.  The initial composition is 100\% H
     (left column), 100\% Fe (middle column), mixed composition
     representative of that observed in Galactic cosmic rays
     (right column).  The acceleration rate is such that
     $\rho_{\rm cut}^{(p)}=10^{20}$~V (top row), $10^{21}$~V
     (middle row), and $10^{22}$~V (bottom row). The IR/Opt/UV
     background is taken to be that at $z$=0, neutrons are
     assumed to decay inside the acceleration zone, $\kappa(\rho)
     \propto \rho^{1/3}$ and $\rho_{\rm max}=10^{25}$~V.  Other
     acceleration parameters are as indicated. Curves are for
     atomic numbers as indicated in the legend, with the topmost
     solid curve being the total spectrum, middle solid curve
     being $Z$=1, and the bottom solid curve being for the total
     neutrino spectrum.}
     \label{CompAfterAccn_A}
\end{figure*}

\begin{figure*}
\centering
   \includegraphics[angle=90,height=14cm,width=19cm]{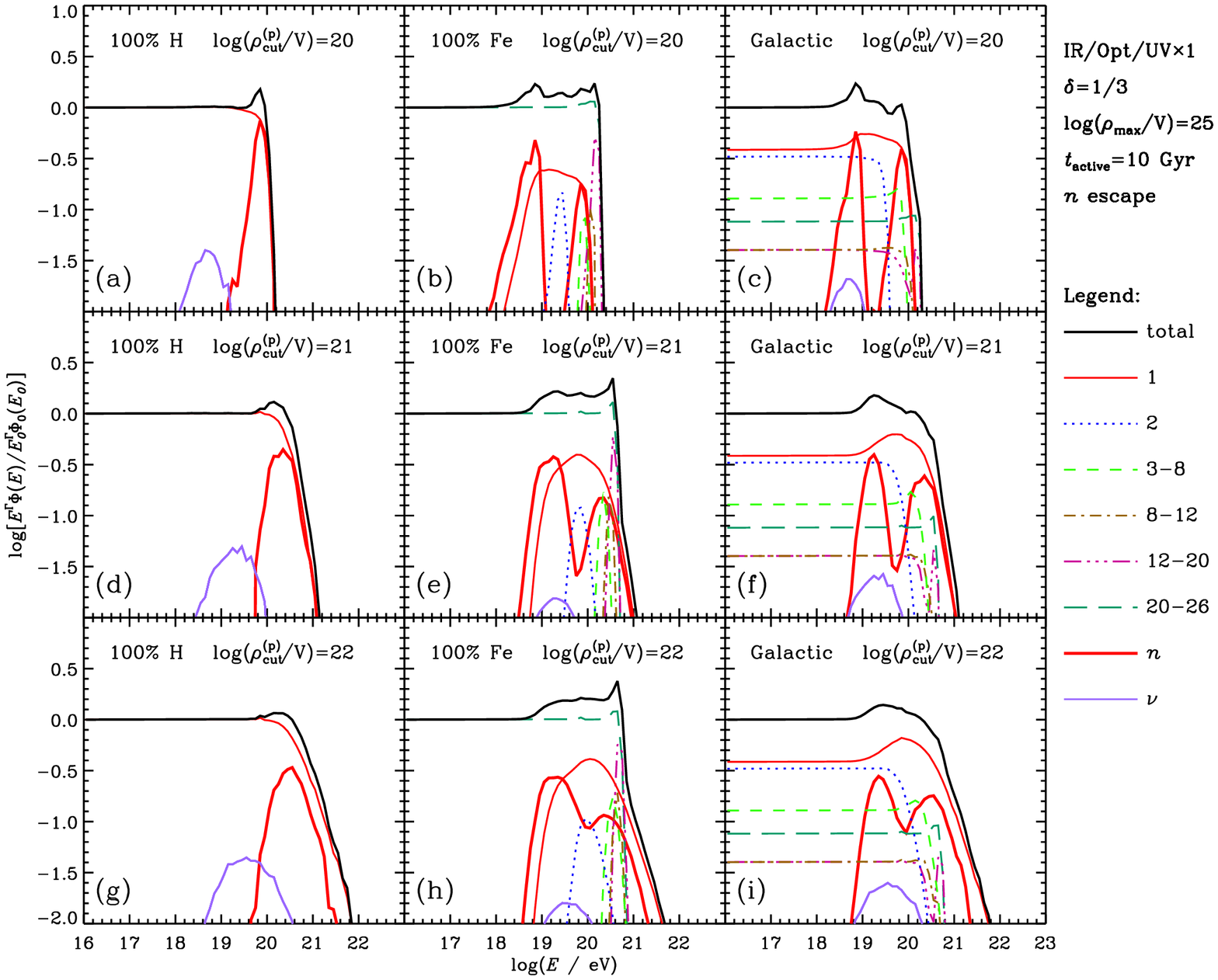}
     \caption{As Fig.~\ref{CompAfterAccn_A} except that neutrons are assumed to rapidly escape from the acceleration zone where they decay into protons. Curves are as in Fig.~\ref{CompAfterAccn_A} except that the additional thick solid curve shows the contribution of neutrons.}
     \label{CompAfterAccn_B}
\end{figure*}

We note that the term $r_{\rm loss}(\rho_{\rm
cut}^{(p)},Z$=1), used to normalize the rigidity dependence of
the acceleration rate, represents the loss rate of a
proton at a its cut-off rigidity $\rho_{\rm cut}^{(p)}$ in the
extragalactic photon background (i.e. the CMB and the inergalactic
IR/opt/UV background). Using the parameter $\rho_{\rm cut}^{(p)}$
(i.e. the rigidity at which the loss rate would equal the
acceleration rate for protons in an accelerator ``bathed'' in the
intergalactic photon background) as a reference will turn out to be
very convenient for estimating the effect of different physical
parameters ($\rho_{\rm max}$, $\delta$ and the scaling factor of
the IR/opt/UV background) on the spectrum of accelerated protons
and nuclei.  Our Monte Carlo results and the conclusions of our
study will be presented in the following two sections.

\section{Results}

\begin{figure*}
\centering
   \includegraphics[angle=90,height=14cm,width=19cm]{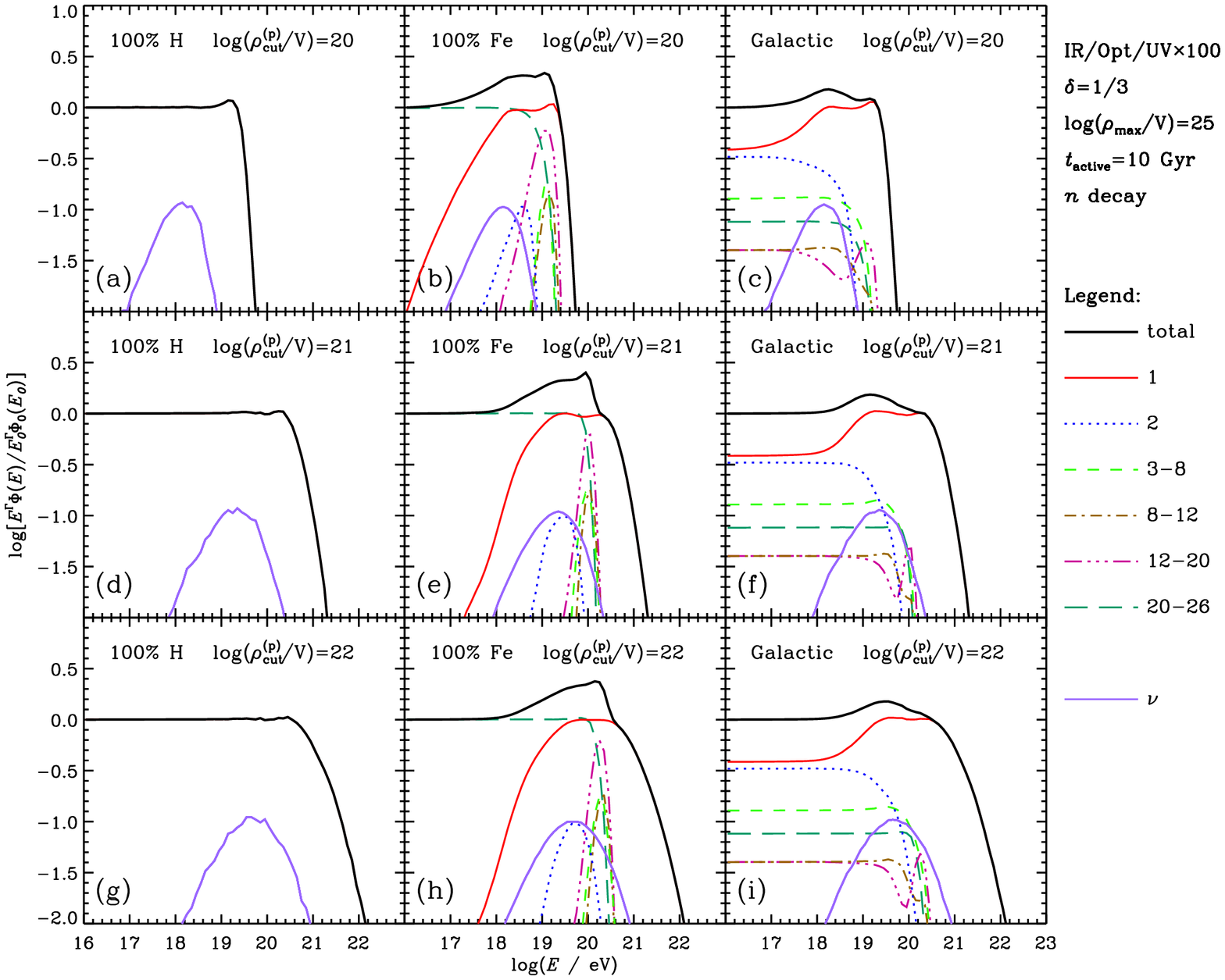}
     \caption{As Fig.~\ref{CompAfterAccn_A} except that the IR/Opt/UV is taken to be 100 times that at $z$=0.  }
     \label{CompAfterAccn_C}
\end{figure*}

\begin{figure*}
\centering
   \includegraphics[angle=90,height=14cm,width=19cm]{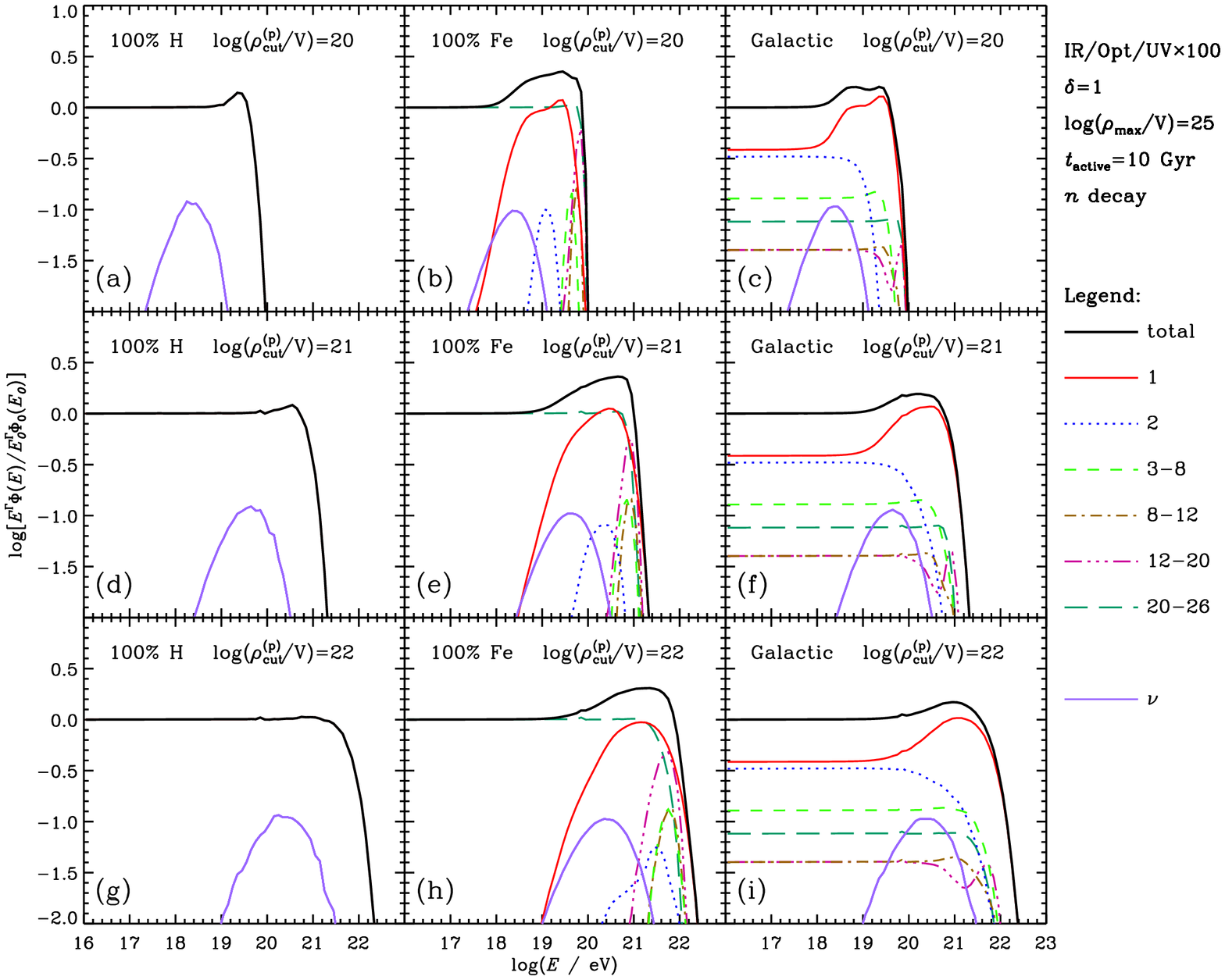}
     \caption{As Fig.~\ref{CompAfterAccn_C} except that $\kappa(\rho) \propto \rho^{1}$. }
     \label{CompAfterAccn_D}
\end{figure*}

In this section, we describe the results of our calculations for
different combinations of the above-mentioned acceleration
parameters.  For each combination of $\delta$, $\rho_{\rm max}$
and the scaling of the IR/opt/UV, we calculate the expected
output spectrum for different values of $\rho_{\rm cut}^{(p)}$
($10^{20}$, $10^{21}$, $10^{22}$ V) and various compositions
injected at the shock: (i) pure proton, (ii) pure iron, and (iii)
a mixed composition similar to low energy Galactic cosmic rays
(see Allard et al., \cite{Denis2005a}, and references therein).
In Figs.~\ref{CompAfterAccn_A}--\ref{CompAfterAccn_E} we give the
spectra of nuclei, showing contributions of different groups of
elements, accelerated such that in the absence of interactions
and losses the spectra would be proportional to $\rho^{-\Gamma}$,
and in the present work we set the nominal spectral index to be
$\Gamma$=2.  The choice of other reasonable values of $\Gamma$
would make only a minor difference to the relative compositions
of accelerated particles, and the relative contributions of
secondary nucleons, photons, and neutrinos would be somewhat
higher (lower) for harder (softer) spectra, but this would not
modify the main conclusions of the present work.  In these
figures we also show the flux of neutrinos plus antineutrinos
(summed over all flavors) produced during acceleration, and for
the case where neutrons are assumed not to decay inside the
acceleration zone we give the contribution to the total escaping
cosmic rays of neutrons produced during pion photoproduction and
photo-disintegration of nuclei.  Note that unless otherwise
specified we assume that the source is active for 10~Gyr, and
that the injection rate does not depend on time.

Fig.~\ref{CompAfterAccn_A} shows the accelerated cosmic-ray
spectra assuming $\rho_{\rm max}=10^{25}$ V (corresponding to an
infinite shock), Kolmogorov turbulence (diffusion coefficient
rigidity-dependence index $\delta=1/3$), a photon background
equivalent to the extragalactic background, and neutrons decaying
inside the acceleration zone.  For this case, the acceleration
process is interaction-limited because $\rho_{\rm max}$ is
extremely high, and we see that there is no obvious scaling of
the maximum energies of the various species with their mass or
charge -- this is consistent with our expectation based on
Figs.~\ref{Loss1} and \ref{Loss2}.  Indeed, although the
interaction threshold for a given photon background will occur at
energies energies more or less proportional to the mass number of
nuclei, such a scaling is not observed for the various cases
studied in Fig.~\ref{CompAfterAccn_A}.  This is because the mean
free paths also scale with the mass, which means that the energy
at which the photo-disintegration mean free path becomes of the
same order as the acceleration length scales in a non-trivial way
with mass number $A$.  However, heavy nuclei can reach higher
energies than light and intermediate nuclei due to their higher
interaction energy threshold and higher acceleration rate at a
given energy.

Once the acceleration rate becomes lower than the interaction
rate for a given component, a sharp cut-off takes place as the
nucleus changes to a lower mass and lower energy. As can be seen
in Figs.~\ref{Loss1} and \ref{Loss2}, the cut-off energy depends
on $\rho_{\rm cut}^{(p)}$ and $\delta$ (see below). Due to the
very low values of the interaction length for nuclei with CMB
photons, such cut-offs seem to be extremely difficult to avoid
even for very rapid acceleration (corresponding to high values of
$\rho_{\rm cut}^{(p)}$).  The proton component also experiences a
cut-off at energies depending on $\rho_{\rm cut}^{(p)}$ because
of the increase of the photopion interactions loss rate with
energy and the subsequent possibility of protons to fall out of
the box, the latter being more important for protons than for
nuclei because nuclei mainly lose energy by ejecting nucleons.
This cut-off is however smoother than in the case of nuclei due
the different nature of the energy losses.  In the case we study
here, i.e.\ acceleration not limited by confinement and a source
active during the Hubble time, protons can reach an energy
slightly higher than heavy nuclei, their lower acceleration rate
(at a given energy) being compensated for by a lower interaction
rate.  When the acceleration rate is high enough (e.g.\
$\rho_{\rm cut}^{(p)} > 10^{21}$ eV) most species of nuclei (even
the lightest) can reach energies above $10^{20}$ eV.

For the case of a mixed composition or 100\% iron there is a bump
in the spectrum at the highest energies due to secondary
nucleons, and it starts quite abruptly at the energy at which the
photodisintegration interaction rate becomes higher than the
acceleration rate.  Secondary protons and protons from decay of
secondary neutrons in the box as assumed
Fig~\ref{CompAfterAccn_A} continue to be accelerated and produce
a tail in the energy spectrum at the highest energies.
 
The alternative case in which neutrons escape before decaying is
equivalent to adding a supplementary escape term as approximately
half of the produced secondary nucleons leave the box without
being further accelerated.  In this case, protons also have an
additional probability of escaping after each pion
photoproduction interaction equivalent to the probability of
isospin flip (see M\"ucke et al., 2000).  As a consequence, the
cut-off in the proton component in Fig.~\ref{CompAfterAccn_B} is
sharper than for the case of neutron decay inside the box shown
in Fig~\ref{CompAfterAccn_A}.  In the 100\% proton case there is
a bump in the spectrum due to neutrons which starts at the energy
at which the pion photoproduction interaction rate becomes higher
than the gain rate for a given value of $\rho_{\rm cut}^{(p)}$.
When nuclei are injected there is also a second bump at lower
energies corresponding to the interaction threshold for
photo-disintegration.

\subsection*{Case of a high ambient photon background}

As mentioned earlier, the photon background in astrophysical
sources could be significantly higher than in intergalactic
space, and to mimic such stronger fields we have scaled the
IR/opt/UV intergalactic background upward by various factors to
investigate the influence of the photon density on the
accelerated spectrum and composition.  We have kept the other
parameters the same as for Fig.~\ref{CompAfterAccn_A}, i.e.\
$\delta=1/3$, $\rho_{\rm max}=10^{25}$~V and neutrons decaying
inside the box, and show in Fig.~\ref{CompAfterAccn_C} the result
for an IR/opt/UV background 100 times higher than in
intergalactic space.  As can be seen in Fig.~\ref{FSRQ}, such a
background is comparable to the background expected in radio
lobes located $\sim$200 kpc from a relatively powerful FSRQ and
so can serve as a typical reference for what can be expected in
some classes of source (see discussion below).  

Figs.~\ref{Loss1} and \ref{Loss2} show that such a high
background will lower the energy at which the spectrum of a
particular species cuts off as the balance between acceleration
and energy losses/photo-disintegration losses occurs at a lower
energy.  We see in Fig.~\ref{CompAfterAccn_C} that this effect is
stronger for nuclei than protons, but that nevertheless most
species can still be accelerated well above $10^{19}$ eV. We have
found that this applies even for IR/opt/UV backgrounds 1000 times
higher than in the extragalactic space, but that when IR/opt/UV
background is $10^4$ times higher than in extragalactic space
protons and nuclei are not accelerated above $10^{19}$ eV
whatever the value of $\rho_{\rm cut}^{(p)}$.

High photon backgrounds present at sources are usually
thought to eliminate nuclei during the acceleration process
giving rise to a 100\% proton composition.  However, we find this
not to be the case for any combination of the acceleration
parameters that we have explored.

\subsection*{Influence of rigidity dependence of the diffusion coefficient}

The energy dependence of the acceleration rate depends on the
rigidity dependence of the diffusion coefficient $\kappa \propto
\rho^\delta$ in the vicinity of the shock.  As can be seen in
Figs.~\ref{Loss1} and \ref{Loss2}, the power-law index $\delta$
affects the competition between acceleration and energy losses.
We show in Fig.~\ref{CompAfterAccn_D} results for $\delta=1$
(Bohm scaling) while keeping all other parameters the same as
used for Fig.~\ref{CompAfterAccn_C} for which $\delta=1/3$
(Kolmogorov turbulence) was used.  We see that as result of the steeper
dependence of the acceleration rate on energy the cut-offs in the
spectra of the various nuclear species are shifted to higher
energies with the gap between protons and nuclei being reduced,
and heavy nuclei being able to reach as high energies as protons
even with the IR/Opt/UV background 100 times that in
intergalactic space.

\begin{figure*}
\centering
   \includegraphics[angle=90,height=14cm,width=19cm]{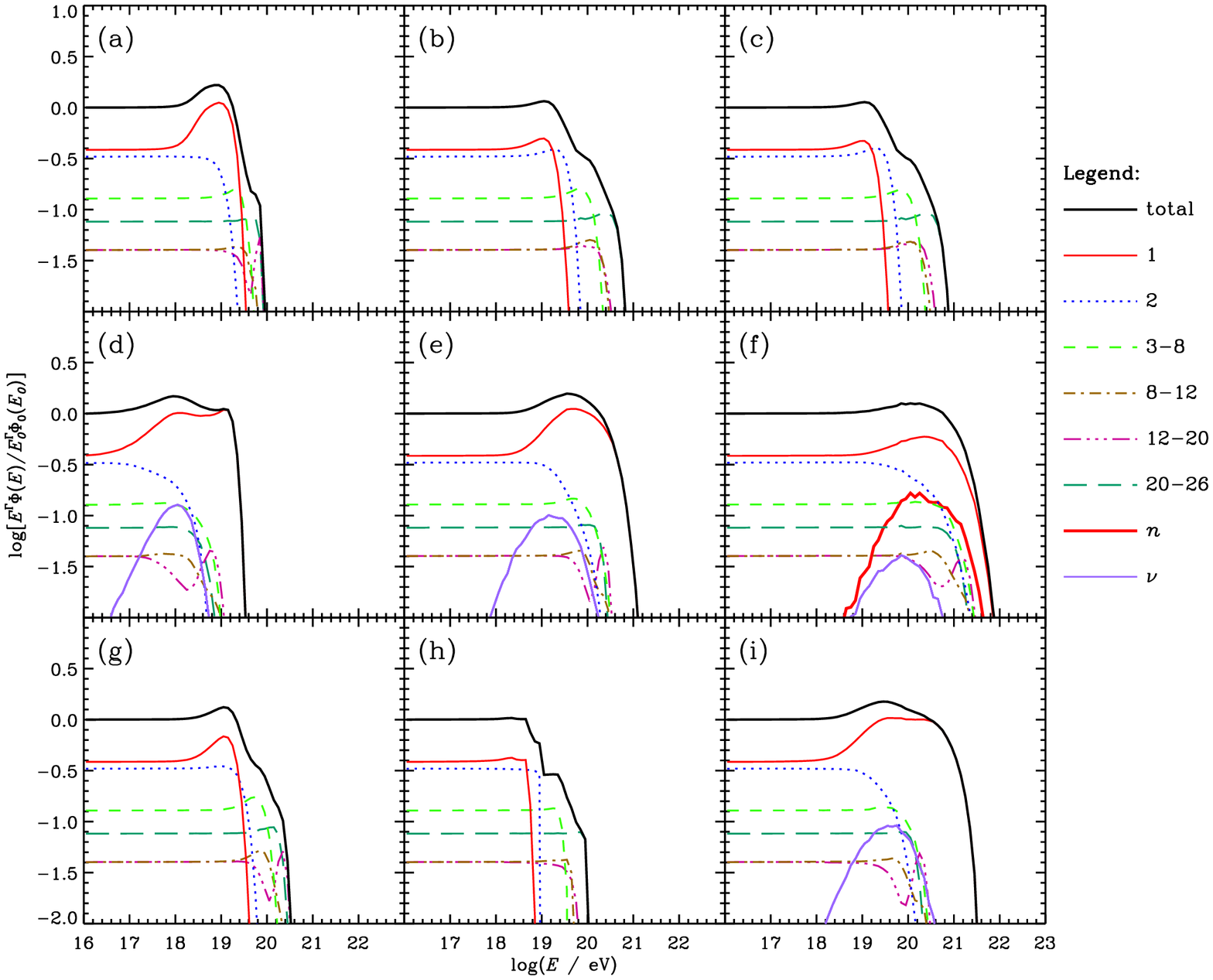}
     \caption{ Miscellaneous combinations of parameters: (a), (b)
and (c) As Fig.~\ref{CompAfterAccn_A} except that  $\delta=1$, $\rho_{\rm
max}=10^{19}$~V and only the mixed composition
is shown. (d) Mixed composition, $\rho_{\rm
cut}^{(p)}=10^{22}$~V, IR/Opt/UV$\times$10000, $\delta=1/3$,
$\rho_{\rm max}=10^{25}$~V and neutrons decay inside the box. (e)
Same as (d) but with $\delta=1$. (f) Mixed composition,
$\rho_{\rm cut}^{(p)}=10^{22}$~V, IR/Opt/UV$\times$1000,
$\delta=1$, $\rho_{\rm max}=10^{25}$~V and neutrons escape from the
box. (g) Same as (c) but with IR/Opt/UV$\times$10000. (h) Same as
Fig.~\ref{CompAfterAccn_C} for the mixed composition and $\rho_{\rm
cut}^{(p)}=10^{22}$~V except that the source activity duration is
10 Myr. (i) Same as (h) but for a source activity duration of 100 Myr.  }
     \label{CompAfterAccn_F}
\end{figure*}

We introduce here Fig.\ref{CompAfterAccn_F} to illustrate some of
the effects and dependencies that we have noted already, and to
highlight them in a form which more clearly shows their origin.  The
effect of changing the rigidity dependence of the acceleration
rate noted above is even more pronounced if neutrons escape from the
acceleration zone (see Fig.~\ref{CompAfterAccn_F}f).  The
comparison between Figs.~\ref{CompAfterAccn_F}d and
\ref{CompAfterAccn_F}e shows there to be a dramatic increase in
the maximum energies of protons and nuclei in a very
high radiation field (a factor $10^4$ increase in IR/Opt/UV) when
the rigidity dependence of the diffusion coefficient is changed
from $\rho^{1/3}$ (Kolmogorov) to $\rho^1$ (Bohm).

Another consequence of a stronger dependence of the acceleration
rate on $\rho$ is a sharper cut-off of the proton component
because the gain rate more rapidly decreases above $\rho_{\rm
cut}^{(p)}$ and it is then very difficult to continue to
accelerate protons once they start to interact, even if they do
not escape from the source immediately.

\begin{figure*}
\centering
   \includegraphics[angle=90,height=14cm,width=19cm]{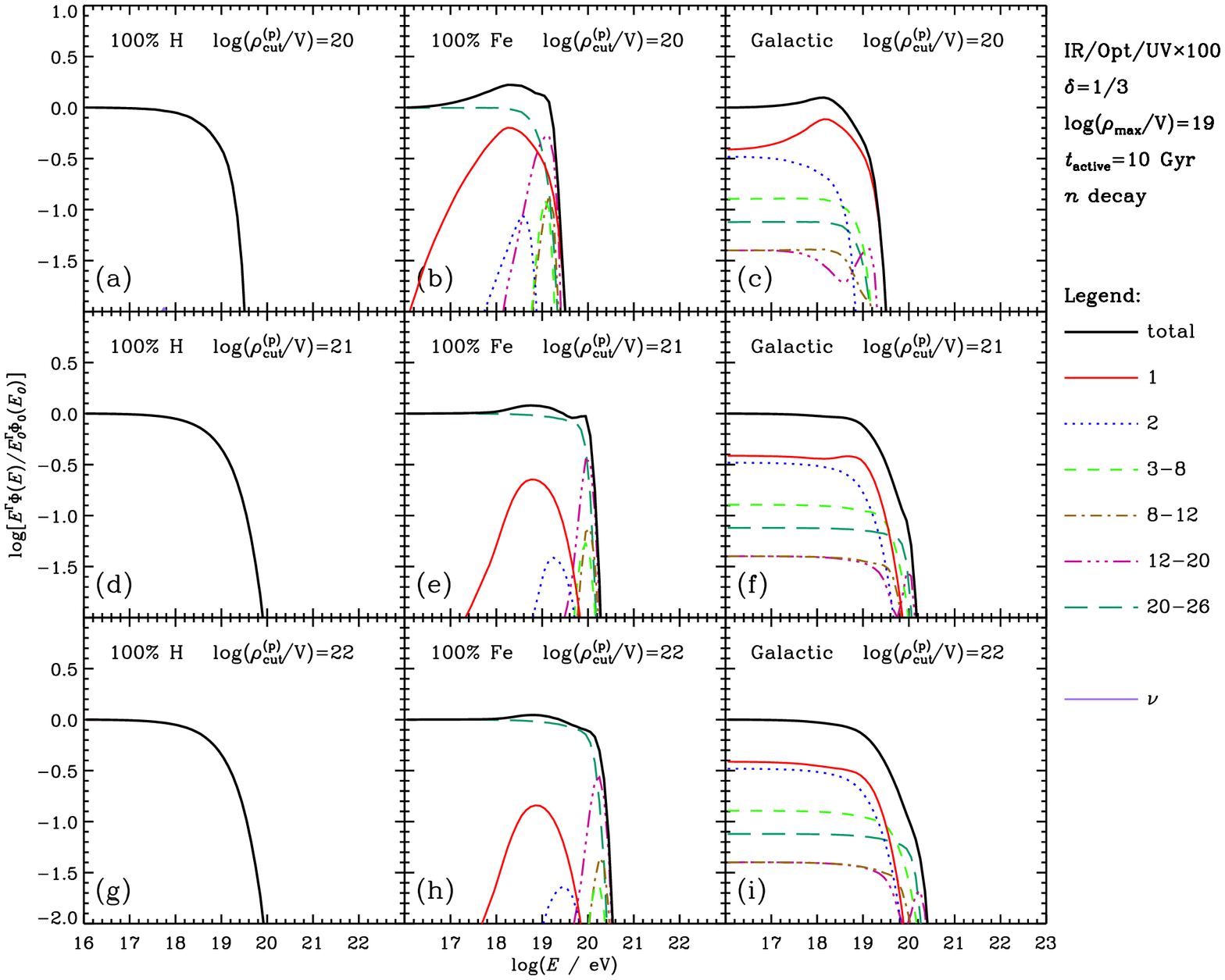}
     \caption{As Fig.~\ref{CompAfterAccn_A} except that $\rho_{\rm max}=10^{19}$~V. }
     \label{CompAfterAccn_E}
\end{figure*}

\subsection*{Confinement-limited acceleration}

So far we have only considered cases where the acceleration is
limited by interactions which we model by setting $\rho_{\rm
max}$ to a very high value, $10^{25}$ V.  However, astrophysical
sources are of finite size and so are expected also to have
limited capability of confining high rigidity cosmic rays, e.g.\
when their larmor radius becomes comparable to the size of the
acceleration medium.  Confinement-limited acceleration processes
can be modeled by lowering the value of $\rho_{\rm max}$.
Fig.~\ref{CompAfterAccn_E} shows the result of our calculations
assuming $\rho_{\rm max}=10^{19}$~V while keeping all the other
parameters the same as in Fig.~\ref{CompAfterAccn_C} (neutrons
decay inside the box, $\delta=1/3$ and a scaling factor of 100
for the IR/opt/UV background).  In all cases, we see that the
proton component cuts off around $10^{19}$ eV due to the
confinement limitation.  With a rigidity cut-off one would expect
that in absence of energy losses the cut-off energies would be
proportional to the charge for the various species, but one can
see in Fig.~\ref{CompAfterAccn_E} that for this choice of the
parameters this scaling only applies very crudely and only for
the rapid acceleration case ($\rho_{\rm cut}^{(p)}=10^{22}$ eV).
The acceleration of nuclei still appears to be
interaction-limited, but nevertheless heavy nuclei are always
accelerated above $10^{20}$ eV when $\rho_{\rm cut}^{(p)}\ge
10^{21}$ eV and are accelerated to higher energies than the
proton component.

We find that for $\rho_{\rm max}=10^{19}$~V provided there is a
steep energy dependence of the rigidity gain rate ($\delta$=1)
the expected scaling of the cut-off energies with $Z$ applies
for the case of a photon background with the IR/Opt/UV 100
times that in intergalactic space (top panel of
Fig.~\ref{CompAfterAccn_F}) and even with the IR/Opt/UV 1000
times higher (Fig.~\ref{CompAfterAccn_F}g).  The expected scaling
also applies for lower photon backgrounds,

For higher values of $\rho_{\rm max}$, e.g. $10^{21}$ V, there
will be very little difference for nuclei to the $\rho_{\rm
max}=10^{25}$~V case (e.g.\ $\rho_{\rm cut}^{(p)}=10^{21}$ V and
$10^{22}$ V in Fig.~\ref{CompAfterAccn_C}) as interactions
already cut off their spectra at rigidities below $Z\times
10^{21}$ eV.  Only the proton spectrum will be
confinement-limited reducing the gap between the
cut-off energies of protons and nuclei.

\subsection*{Influence of duration of source activity}

Some of the astrophysical sources that are candidates for the
acceleration of cosmic-rays to the highest energies, e.g.\ FSRQ
which according to unification schemes (e.g.\ Urry \& Padovani
(\cite{PadovaniUrry1992}) are the beamed counterparts of the
powerful Fanaroff-Riley Class II (FR-II) radio galaxies, are
transitory sources with a life time small compared to the age of
the universe. The finite duration of the source activity (by
activity we mean the period during which the source is able to
accelerate cosmic-rays) does not necessarily impact on the
spectrum and composition of accelerated cosmic rays. Indeed, if
the acceleration time scale is small compared to the duration of
source activity, there will essentially be no difference. An
example is shown in Fig.~\ref{CompAfterAccn_F}i, where the source
parameters of Fig.~\ref{CompAfterAccn_C} are used except for the
source activity duration which is only 100 Myr. We see that for a
high acceleration rate ($\rho_{\rm cut}^{(p)}=10^{22}$~V) the
output is essentially identical to that shown in
Fig.~\ref{CompAfterAccn_C}, with the nuclear components being
almost indistinguishable between the two cases and the only major
difference being that the high energy tail of the proton
component is absent when the source has the shorter lifetime.

The situation changes when the source activity duration is less
than the time needed to accelerate protons at the highest
energies. Fig.~\ref{CompAfterAccn_F}h shows the spectrum and
composition for the same parameters as
Fig.~\ref{CompAfterAccn_F}i except that the source activity
duration is only 10 Myr. In this case, the proton component is
even more affected due to its lower acceleration rate and the
maximum energies scale proportionally to the charge of the
species.  In fact the proton cut-off rigidity, at $6\times
10^{18}$~V, is significantly lower than $10^{20}$~V which would
naively be expected based on reading off $\rho$ in
Fig.~\ref{Loss1} at $x$=3~Mpc (corresponding to 10~Myr) for the
$\rho_{\rm cut}^{(p)}=10^{22}$~V and $x_{\rm gain}\propto
\rho^{1/3}$ case.  The reason for this is that for injection of a
proton at $\rho=\rho_0\ll \rho_{\rm cut}^{(p)}$ at time $t$=0,
and acceleration for time $t=x/c$, the proton has a rigidity of
$\rho_{\rm max}\approx [x\delta /x_{\rm gain}(\rho_{\rm
cut}^{(p)})]^{1/\delta}\rho_{\rm cut}^{(p)}$.  Once more, the
heaviest components are close to the default case of
Fig.~\ref{CompAfterAccn_C} because of their higher charge (and
acceleration rate), even for such a short acceleration time.

\section{Conclusions}

We have studied DSA of protons and nuclei in astrophysical
sources in the presence of background radiation fields and there
are some general conclusions we can draw.  Provided the ambient
magnetic field is not too high (see Fig.~\ref{fig_rho_cut_vs_b}
for protons) we find that when the acceleration mechanism is not
limited by confinement protons usually reach higher energies than
nuclei which are more limited by interactions.  This conclusion
is particularly true for the conservative assumption that the
turbulence in the vicinity of the shock is of Kolmogorov-type
(i.e.\ a slow dependence of the acceleration rate with energy).
If the turbulence brings the diffusion coefficient dependence
closer to the Bohm regime we find that heavy nuclei can usually
reach maximum energies as high as those of protons, even in the
case of very high ambient photon fields.  Whenever the
acceleration mechanism is not confinement-limited we find there
to be no trivial scaling of the maximum energies of the various
species with their mass or atomic number.

It is beyond the scope of the present paper to calculate the
expected spectra at Earth for each source model, and we refer the
reader to Allard et al.\ (\cite{Denis2005a}-\cite{Denis2008}) for
detailed studies of the propagation of cosmic-ray nuclei in the
extragalactic radiation fields.  However, the non-trivial scaling
of the maximum energy with the mass or atomic number should not
affect significantly the results obtained in propagation studies
in which maximum energies are usually assumed to be proportional
to the charge.  Indeed, we have seen that secondary nucleons
produced in interactions within the accelerator continue to be
accelerated (except for escaping neutrons) and can then reach
higher energies.  On the other hand, nuclei at energies above
$\sim A \times (4\!\times\!10^{18}$ eV) are expected to be almost
fully photo-disintegrated by CMB photons during propagation (see
Allard et al., \cite{Denis2008}), and this means that nuclei
accelerated above these energies would not be seen at Earth and
would simply release their nucleons at energies a factor $A$
lower.  Whether or not nuclei above $\sim A
\times4\!\times\!10^{18}$ eV are photo-disintegrated within the
source or during the propagation will make little difference to
what is observed at Earth.  Furthermore, any bumps in the
spectrum due to secondary nucleons should be smoothed by source
to source variability as well as by cosmological effects during
the propagation.  Bumps in the cosmic ray spectrum would remain
after propagation of cosmic-rays from a single source, but not
remain after summing the contributions of all sources in the
Universe because of the large range of propagation times from the
sources to the earth except for the case of a dominant
contribution by a very nearby source.

One important point to emphasize is that even in the presence of
strong photon backgrounds, nuclei are accelerated to energies
above $10^{19}$ eV whenever protons are, even though nuclei are
more limited by interactions than nucleons.  Hence, our results
show that DSA at sources with physical conditions that would
cause nuclei to be fully photodisintegrated during acceleration
but allow protons to reach energies above $10^{20}$ eV are
excluded.  In other words a pure proton composition is unlikely
to result from an acceleration process whenever a mixed
composition is injected at a shock.  A 100\% proton composition
would therefore require a different explanation such as injection
of only protons at the shock, or propagation in a medium
optically thick to photodisintegration of nuclei in the vicinity
of the source.

It could be argued that our modeling of the photon background in
which we scaled the extragalactic IR/Opt/UV by factors between 1
and 10000 is too simplistic.  However, we shall now consider our
results briefly in the context of acceleration in the radiation
field at a shock at a hot-spot of an FR-II radio galaxy.  FR-II
radio galaxies are thought to be the UN-beamed counterparts of
FSRQ.  Given the spectral energy distribution (SED) of a
``typical'' FSRQ, we may infer the radiation field at the likely
diffusive shock acceleration sites at the far end of an FR-II
jet.  In Fig.~\ref{FSRQ} we showed the radiation field expected
at 200~kpc along the jet from 0208-512 based on the SED for this
FSRQ given in fig.~5 of Tavecchio et al.\ (\cite{Tavecchio2002}),
and compared it with the IR/Opt/UV at $z$=0, and IR/Opt/UV fields
that are 100 and 1000 times higher.  We see that while the low
energy part of the SED is important, the high energy part of the
SED, i.e.\ X-rays and above, will have little effect as target
photons for pion photoproduction or photo-disintegration of
nuclei.  However, there will actually be a very large range in
the normalizations of the FSRQ blazar field at the hot-spot for
two reasons.  First, the distance from the core to the hot-spots
for FR-II is probably in the range 50--800 kpc, giving rise to a
range of a factor of about 300 in radiation density.  Secondly,
there is also a large range in luminosity of FSRQ -- Zhou et
al. (\cite{Zhou2007}) found that $\log[L_\nu\mbox{(1.5 GHz) / W
Hz$^{-1}$}]$ to range from 26.4 to 29.0 with a mean value 27.7
for 45 FSRQ.  This is to be compared with that of 0208-512 for
which $\log[L_\nu\mbox{(1.5 GHz)]} \approx 28.5$ W
Hz$^{-1}$. Nevertheless, as the shape of the SED of the
extragalactic of photon background is not very different from the
FSRQ SED shown in Fig.~\ref{FSRQ} our results can then give some
hint of the influence of the photon backgrounds in this type of
source.  However, we plan to consider more specifically the
acceleration of cosmic-rays in FR-II galaxies in a future
paper.

Interaction-limited acceleration processes are also expected to
result in production of gamma-rays and neutrinos. As shown by
Protheroe (\cite{Protheroe2004}), the position of the neutrino peak can help to
constrain the acceleration parameters in proton acceleration
sources and the characteristics of the source. However, as can be
seen in the various figures in Sect.~4, when we include nuclei
neutrino fluxes cannot give any strong constraints on the
injected composition at the shocks.

When the size of the source becomes comparable to or smaller than
the Larmor radius of protons at the highest energies, the
acceleration becomes limited by confinement, in which case heavy
nuclei are accelerated to higher energies than protons unless the
acceleration is very slow and/or the photon background is very
high.  In the case of the maximum proton energy being below
$10^{20}$ eV the maximum energies of all species are proportional
to their charge, even in the presence of high photon backgrounds,
provided the acceleration parameters are favorable ($\rho_{\rm
cut}^{(p)}\geq10^{21}$ eV and $\delta$ close to ``Bohm
scaling'').  Such a composition is especially interesting in the
context of recent cosmic-ray data.  Indeed, preliminary results
from the Pierre Auger Collaboration show the composition possibly
getting heavier above $10^{19}$ eV (see Unger et al., 2007).  If
this trend is confirmed, the most likely explanation would
involve confinement-limited acceleration with the maximum energy
of the species scaling with the charge.  A composition getting
heavier above $10^{19}$ eV is indeed very difficult to obtain
with propagation effects only (see Allard et al.,
\cite{Denis2005a}-\cite{Denis2008}) when protons are accelerated
up to the highest energies.  We have also seen in previous
sections that in most cases if the acceleration is
interaction-limited then nuclei are not accelerated to higher
energies than protons.  Interestingly, it has been shown recently
(Allard et al., \cite{Denis2008}) that the highest energy
cosmic-ray spectrum can be successfully fitted by assuming a low
proton maximum energy ($\sim 10^{19}$ eV) and a charge-scaling
maximum energy for the other species, providing abundances of
heavy nuclei slightly higher than in the Galactic cosmic rays.
Although other possibilities are not ruled out, such a scenario
where protons are not required to be accelerated above $10^{20}$
eV but where the highest energy cosmic-rays are provided by the
heavy component would completely change the expectations and
constraints on possible accelerators like blazars (M\"{u}cke et
al., 2003) or radio-galaxies.  We shall investigate these
implications in forthcoming papers. Finally, we note that
confinement-limited acceleration is unlikely to provide strong
neutrino fluxes at high energy.

\begin{acknowledgements}
      This work was supported in part by the Australian Research
      Council through Discovery Project grant DP0881006 and by
      the Australia-France Cooperation Fund in Astronomy.
\end{acknowledgements}

\end{document}